%% file: Hplus_taunu-PAPER.tex
\newcommand*{\ATLASLATEXPATH}{}
\documentclass[cernpreprint,texlive=2011,txfonts,UKenglish]{\ATLASLATEXPATH atlasdoc}

\usepackage{bm}
\usepackage{units}
\usepackage{hyperref}
\usepackage{placeins}
\usepackage{multirow}

\usepackage{\ATLASLATEXPATH atlaspackage}
\usepackage{\ATLASLATEXPATH atlasbiblatex}

\usepackage{\ATLASLATEXPATH atlascontribute}

\usepackage{\ATLASLATEXPATH atlasphysics}

\graphicspath{{logos/}{}}

\usepackage{atlascover}
\usepackage{atlasphysics}

\AtlasTitle{Search for charged Higgs bosons produced in association with a top quark and decaying via $\Hpm \to \tau\nu$ using $pp$ collision data recorded at $\sqrt{s} = \unit[13]{\TeV}$ by the ATLAS detector}

\author{The ATLAS Collaboration}

\AtlasRefCode{HIGG-2015-11}

\PreprintIdNumber{CERN-PH-2016-056}

\AtlasAbstract{Charged Higgs bosons produced in association with a single top quark 
and decaying via $\Hpm \to \tau\nu$ are searched for with the ATLAS experiment at 
the LHC, using proton--proton collision data at $\rts = \unit[13]{\TeV}$ corresponding to 
an integrated luminosity of $\unit[3.2]{\ifb}$. The final state is characterised by 
the presence of a hadronic $\tau$ decay and missing transverse momentum, as well 
as a hadronically decaying top quark, resulting in the absence of high-transverse-momentum 
electrons and muons. The data are found to be consistent 
with the expected background from Standard Model processes. A statistical 
analysis leads to 95\% confidence-level upper limits on the production 
cross section times branching fraction, $\sigma(pp \to [b]t\Hpm) \times 
\mbox{BR}(\Hpm \to \tau\nu)$, between 1.9~pb and 15~fb, for charged Higgs 
boson masses ranging from 200 to 2000~\GeV. The exclusion limits for this 
search surpass those obtained with the proton--proton collision data 
recorded at $\rts = \unit[8]{\TeV}$.}

\AtlasJournalRef{Phys. Lett. B 759 (2016) 555-574}
\AtlasDOI{10.1016/j.physletb.2016.06.017}

\AtlasCoverSupportingNote{Search for charged Higgs bosons in the $\tau$+jets final state with $\unit[3.2]{\ifb}$ of $pp$ collision data recorded at $\rts = \unit[13]{\TeV}$ with the ATLAS experiment}{https://cds.cern.ch/record/2051488}

\AtlasCoverCommentsDeadline{XX March 2016}

\AtlasCoverAnalysisTeam{Br{\"u}ckman de Renstrom, Pawel; Carrillo-Montoya, German; Feremenga, Last; Ferrari, Arnaud; Griffiths, Justin; Gross, Eilam; Hadavand, Haleh Khani; Kaczmarska, Anna; Kowalewska, Anna Bozena; Pitt, Michael; Rangel-Smith, Camila; Schaarschmidt, Jana; Yu, Jaehoon; {\"O}hman, Henrik}

\AtlasCoverEdBoardMember{Coadou, Yann (chair); Sekula, Stephen; Trigger, Isabel} 

\AtlasCoverEgroupEditors{atlas-HIGG-2015-11-editors@cern.ch}

\AtlasCoverEgroupEdBoard{atlas-higg-2015-11-editorial-board@cern.ch}

\hypersetup{pdftitle={ATLAS},pdfauthor={The ATLAS Collaboration}}

\begin{document}

\maketitle

\section{Introduction}
\label{sec:Introduction}
\input{introduction.tex}

\section{Data and simulated events}
\label{sec:Samples}
\input{samples.tex}

\section{Object reconstruction and identification}
\label{sec:PhysicsObjects}
\input{physics_objects.tex}

\section{Event selection and background modelling}
\label{sec:Analysis}
\input{Selection.tex}

\section{Systematic uncertainties}
\label{sec:syst}
\input{Systematics.tex}

\section{Results}
\label{sec:Results}
\input{Results}

\FloatBarrier
\section{Conclusion}
\label{sec:Conclusion}

\input{conclusion.tex}

\section*{Acknowledgements}
\input{Acknowledgements.tex}

\printbibliography
\newpage 
\input{atlas_authlist}

\end{document}

%% file: introduction.tex
Following the discovery of a neutral scalar particle at the Large Hadron Collider 
(LHC) in 2012~\cite{Aad:2012tfa,Chatrchyan:2012ufa}, an important question is 
whether this new particle is the Higgs boson of the Standard Model (SM) or part 
of an extended Higgs sector. Charged Higgs bosons\footnote{In the following, 
charged Higgs bosons are denoted $\Hp$, with the charge-conjugate $\Hm$ always 
implied. Similarly, generic symbols are used for their decay products.} 
appear in several non-minimal scalar sectors, where a second doublet~\cite{2hdm} or 
triplets~\cite{trip1,trip2,trip3,trip4,trip5} are added to the SM Higgs doublet. 
The observation of a charged Higgs boson would therefore clearly indicate physics 
beyond the SM.

The ATLAS and CMS collaborations have searched for light charged Higgs bosons, produced in 
top-quark decays, using proton--proton ($pp$) collisions at $\rts$ = 7--8 \TeV\ 
in the $\tau\nu$~\cite{Aad:2012tj,Aad:2012lea,taunupaper,:2012cwb,PAS-HIG-14-023} 
and $cs$~\cite{cs-ATLAS,cs-CMS} decay modes. Using data collected at $\rts = \unit[8]{\TeV}$, 
charged Higgs bosons heavier than the top quark were also searched for, using final states 
originating from both the $\tau\nu$ and $tb$ decay modes~\cite{taunupaper,tbpaper,PAS-HIG-14-023}. 
Vector-boson-fusion $H^{+}$ production was also searched for by ATLAS using the $WZ$ final 
state~\cite{WZpaper}. No evidence of a charged Higgs boson was found in any of these searches.

\begin{figure}[h!]
\begin{center}
\includegraphics[height=4cm]{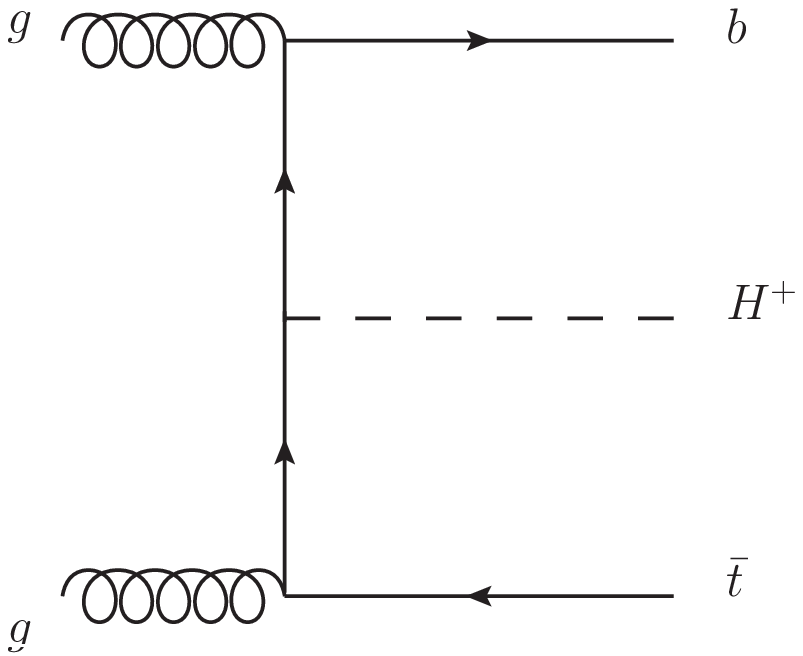}
\hspace*{1cm}
\includegraphics[height=4cm]{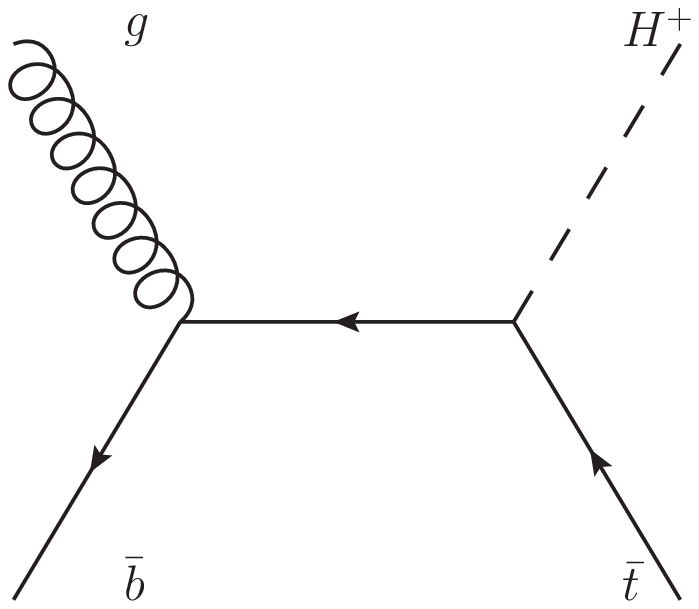}
\end{center}
\caption{Leading-order Feynman diagrams for the production of a charged Higgs boson with
a mass $m_{\Hp}>m_{\text{top}}$, in association with a single top quark (left in the 4FS, and 
right in the 5FS). \label{fig:hplus_feynman_production}}
\end{figure}

For $m_{\Hp}$ greater than the top-quark mass $m_{\text{top}}$, which is the 
mass range of interest in this paper, the main production mode of a charged 
Higgs boson at the LHC is expected to be in association with a top 
quark~\cite{tHplus-ref1,tHplus-ref2,tHplus-ref3}. The corresponding Feynman 
diagrams are shown in Figure~\ref{fig:hplus_feynman_production}. When calculating 
the corresponding cross section in a four-flavour scheme (4FS), 
$b$-quarks are dynamically produced, whereas in a five-flavour scheme (5FS), 
the $b$-quark is also considered as an active flavour in the proton. For 
model-dependent interpretations, 4FS and 5FS cross sections are averaged 
according to Ref.~\cite{Harlander:2011aa}. In two-Higgs-doublet models (2HDMs), the 
production and decay of the charged Higgs boson also depend on the parameter $\tanb$, 
defined as the ratio of the vacuum expectation values of the two Higgs doublets, and the 
mixing angle $\alpha$ between the CP-even Higgs bosons. In the alignment limit, where 
$\cos(\beta - \alpha) \simeq 0$, the decay $\Hp \to \tau\nu$ can have a substantial 
branching fraction. In a type-II 2HDM, even when the decay $H^+ \to tb$ dominates, the branching 
fraction BR($\Hp \to \tau\nu$) can reach 10--15\% at large values of $\tanb$~\cite{xsec2}.

This paper describes a search for charged Higgs bosons in $pp$ collisions at 
$\rts = \unit[13]{\TeV}$ using the ATLAS experiment. The production of a 
charged Higgs boson in association with a single top quark and its decay 
via $\Hp \to \tau\nu$ are explored in the mass range 200 to 2000~\GeV, 
extending by 1000~\GeV\ the mass range considered by the ATLAS collaboration 
during Run 1 of the LHC. The final state is characterised by the presence of 
a hadronic $\tau$ decay and missing transverse momentum arising from the $\Hp$ 
decay, as well as a fully hadronic top-quark decay, resulting in the absence of 
high-transverse-momentum electrons and muons. 
The SM prediction is compared to the data, and results for the signal cross section 
times branching fraction $\sigma(pp \to [b]t\Hpm) \times \mbox{BR}(\Hpm \to \tau\nu)$ 
are presented, together with an interpretation in the hMSSM benchmark 
scenario~\cite{hMSSM,hMSSMbis}, in which the light CP-even Higgs boson 
mass $m_h$ is set to 125~\GeV, without choosing explicitly the 
soft-supersymmetry-breaking parameters.

%% file: samples.tex
The ATLAS experiment~\cite{Aad:2008zzm} consists of an inner detector with 
coverage in pseudorapidity\footnote{ATLAS uses a right-handed coordinate system with its origin at the nominal interaction point (IP) in the centre of the detector and the $z$-axis along the beam pipe. The $x$-axis points from the IP to the centre of the LHC ring, and the $y$-axis points upward. Cylindrical coordinates $(r,\phi)$ are used in the transverse plane, $\phi$ being the azimuthal angle around the $z$-axis. The pseudorapidity is defined in terms of the polar angle $\theta$ as $\eta=-\ln\tan(\theta/2)$.}
up to $|\eta|=2.5$, surrounded by a thin 2~T superconducting solenoid,
a calorimeter system extending up to $|\eta|=4.9$ and a muon spectrometer
extending up to $|\eta|=2.7$ that measures the deflection of muon
trajectories in the field of three superconducting toroid magnets. 
The innermost pixel layer, the insertable B-layer (IBL), was added between 
the first and second runs of the LHC, around a new, narrower and thinner 
beam pipe~\cite{ibl}.
A two-level trigger system is used to select events of interest~\cite{ATL-DAQ-PUB-2016-001}.
The integrated luminosity, considering only the data-taking periods 
of 2015 in which all relevant detector subsystems were operational, 
is $\unit[3.2]{\ifb}$ and has an uncertainty of~5\%. It is derived 
following a methodology similar to that detailed in Ref.~\cite{lumi}, 
from a calibration of the luminosity scale using $x$--$y$ 
beam-separation scans performed in August 2015.

Simulated events of $H^+$ production in association with a single top quark are generated 
in the 4FS at the next-to-leading order (NLO) with {\sc MadGraph5\_aMC@NLO}~v.2.2.2~\cite{Alwall:2011uj}
using the NNPDF23LO~\cite{Ball:2012cx} parton distribution function (PDF) set, interfaced to
{\sc Pythia}~v8.186~\cite{Sjostrand:2007gs} with the A14 set of tuned parameters 
(tune)~\cite{ATL-PHYS-PUB-2014-021} for the underlying event.

The main backgrounds are the production of $t\bar{t}$ pairs, 
single top quarks, $W$+jets, $Z/\gamma^{\ast}$+jets and electroweak gauge boson 
pairs ($WW/WZ/ZZ$), as well as multi-jet events. 
For the generation of $t\bar{t}$ pairs and single top quarks in the $Wt$- and $s$-channels, the 
{\sc Powheg-Box}~v2~\cite{powheg,powbox2} generator with the CT10~\cite{ct10one,ct10two} PDF set 
in the matrix-element calculations is used. Electroweak $t$-channel single-top-quark events are 
generated using the {\sc Powheg-Box}~v1 generator. This generator uses the 4FS for the NLO 
matrix-element calculations together with the fixed four-flavour PDF set CT10F4. For this 
process, the top quark is decayed using {\sc MadSpin}~\cite{Artoisenet:2012st}, thereby 
preserving all spin correlations. For all backgrounds above, the parton shower, the 
fragmentation and the underlying event are simulated using {\sc Pythia}~v6.428~\cite{Pythia6} 
with the CTEQ6L1~\cite{cteq} PDF set and the corresponding Perugia 2012 (P2012) tune~\cite{p2012}. 
The top-quark mass is set to 172.5~\GeV\ for all relevant background and signal samples.
The $t\bar{t}$ cross section is calculated at next-to-next-to-leading order (NNLO), 
including soft-gluon resummation to the next-to-next-to-leading logarithmic (NNLL) order, 
with {\sc Top}++ v2.0~\cite{Cacciari:2011hy,Beneke,Barnreuther,Czakon1,Czakon2,Czakon3,Czakon:2011xx}. 
The single-top-quark samples are normalised to the approximate NNLO cross 
sections~\cite{stopxs1,stopxs2,stopxs3}. 
Events containing a $W$ or $Z$ boson with associated jets are simulated using 
{\sc MadGraph5\_aMC@NLO}~v.2.2.2 at LO with the NNPDF23LO PDF set, interfaced to
{\sc Pythia}~v8.186 with the A14 underlying-event tune. These samples are 
normalised to the NNLO cross sections calculated with {\sc FEWZ}~\cite{wz1,wz2,wz3}. 
Finally, diboson processes are simulated using the {\sc Powheg-Box}~v2 generator 
interfaced to the {\sc Pythia}~v8.186 parton shower model. The CT10 NLO set is used 
as the PDF for the hard-scatter process, while the CTEQ6L1 PDF set is used for 
the parton shower. The non-perturbative effects are modelled using the AZNLO 
tune~\cite{AZNLO:2014}. The diboson samples are normalised to their NLO cross 
sections, as computed by the event generator.

Whenever applicable, {\sc Photos++}~v3.52~\cite{photospp} is employed for photon radiation 
from charged leptons, and {\sc EvtGen}~v1.2.0~\cite{EvtGen} is used to simulate
$b$- and $c$-hadron decays. Multiple overlaid $pp$ collisions (pile-up, 
with 14 collisions per bunch-crossing on average) are simulated with the soft QCD processes 
of {\sc Pythia}~v8.186 using the MSTW2008LO~\cite{Martin:2009iq,Martin:2009bu,Martin:2010db} 
PDF set and the A2 underlying-event tune~\cite{ATLAS:2012uec}. All simulated signal and 
background samples are processed through a simulation~\cite{simulation} of the detector 
geometry and response using {\sc Geant4}~\cite{AGO-0301}. Finally, they are processed 
through the same reconstruction software as the data.

In the following, the backgrounds are categorised based on the type of reconstructed objects 
identified as the visible decay products\footnote{This refers to all $\tau$ decay products except the neutrinos.}
of the hadronically decaying $\tau$ candidate ($\tau_{\text{had-vis}}$). Only simulated events 
having a true hadronically decaying $\tau$ at generator level ($\tau_{\text{had}}$) or with 
a charged lepton (electron or muon) misidentified as a $\tau_{\text{had-vis}}$ are kept. Backgrounds 
arising from a jet misidentified as a $\tau_{\text{had-vis}}$ are estimated with a data-driven method.

%% file: physics_objects.tex
In the ATLAS experiment, hadronic jets are reconstructed from energy deposits in the calorimeters, 
using the anti-$k_t$ algorithm~\cite{Cacciari:2008gp, Cacciari:2005hq} with a radius parameter 
$R = 0.4$. In the following, jets are required to have a transverse momentum $\pt > \unit[25]{\GeV}$ 
and $|\eta| < 2.5$. A multi-variate technique (Jet Vertex Tagger) relying on jet energy 
and tracking variables to determine the likelihood that a given jet originates from 
pile-up~\cite{ATLAS-CONF-2014-018} is applied to jets with $\pt < \unit[50]{\GeV}$ 
and $|\eta |<2.4$. Jets arising from $b$-hadron decays are identified by using 
an algorithm that combines impact parameter information with the explicit 
identification of secondary and tertiary vertices within the jet into a 
$b$-tagging score~\cite{bib:AdvancedTaggers,bjet-new}. The minimal requirement 
imposed on the $b$-tagging score in this analysis corresponds to a 70\% efficiency 
to tag a $b$-quark-initiated jet in $t\bar{t}$ events, with rejection rates of 
400 for light-quark-initiated jets, 27 for $\tau_{\text{had}}$-initiated jets and 8 
for $c$-quark-initiated jets, enhanced with respect to Run 1 by the use of the IBL 
and an improved algorithm. 
The tagging efficiencies from simulation are corrected based on the results of 
flavour-tagging calibrations performed with data~\cite{btag-data}. 
 
Candidates for identification as $\tau_{\text{had-vis}}$ arise from jets that 
have $\pt > \unit[10]{\GeV}$ and for which one or three charged-particle tracks are found 
within a cone of size\footnote{$\Delta R =\sqrt{(\Delta \eta)^2+ (\Delta\phi)^2}$, 
where $\Delta \eta$ and $\Delta\phi$ are differences in pseudorapidity and azimuthal 
angle, respectively.} $\Delta R = 0.2$ around the axis of the $\tau_{\text{had-vis}}$ 
candidate~\cite{tau1,tau2}. These objects are further required to have a visible transverse 
momentum ($\pt^\tau$) of at least 40~\GeV\ and to be within $|\eta| < 2.3$. The output of 
boosted decision tree (BDT) algorithms~\cite{tmva} is used in order to distinguish $\tau_{\text{had-vis}}$ 
candidates from jets not initiated by hadronically decaying $\tau$-leptons. This is done separately 
for decays with one or three charged-particle tracks, and for varying values of the identification 
efficiency. In this analysis, a working point corresponding to a 55\% (40\%) efficiency for 
the identification of 1-prong (3-prong) $\tau_{\text{had-vis}}$ objects is used, with 
rejection rates of $\mathcal{O}(10^2)$ for jets.

In this analysis, events with isolated electron or muon candidates with a transverse energy 
or momentum above 20~\GeV\ are rejected. Electron candidates~\cite{ElectronPerformance} are 
reconstructed from energy deposits (clusters) in the electromagnetic calorimeter, associated 
with a reconstructed track in the inner detector. The pseudorapidity range for the
electromagnetic clusters covers the fiducial volume of the detector, $|\eta|<2.47$
(the transition region between the barrel and end-cap calorimeters, $1.37<|\eta|<1.52$,
is excluded). Quality requirements on the electromagnetic clusters and the tracks, as
well as isolation requirements in a cone around the electron candidate based on its
transverse energy and the tracking information, are then applied in order to reduce
contamination from jets. The muon candidates are reconstructed from track 
segments in the muon spectrometer, and matched with tracks found in the inner detector 
within $|\eta|<2.5$~\cite{muons}. The final muon candidates are refitted using 
the complete track information from both detector systems. They must fulfil quality 
requirements including a $\pt$-dependent track-based isolation requirement in a cone of 
variable size around the muon, which has good performance under high pile-up conditions 
and/or when a muon is close to a jet.

When objects overlap geometrically, the following procedures are applied, in this 
order. Every $\tau_{\text{had-vis}}$ candidate that overlaps with a loosely identified 
electron or muon, within a cone of size $\Delta R$ of 0.4 or 0.2, respectively, is 
removed. Then, reconstructed jets are discarded if an electron or a $\tau_{\mathrm{had-vis}}$ 
candidate fulfilling the selection criteria above is found within a cone of size $\Delta R = 0.2$. 

The magnitude $\met$ of the missing transverse momentum~\cite{etmiss1} is reconstructed 
from the negative vector sum of transverse momenta of reconstructed and fully calibrated 
objects (collected in the hard term), as well as from reconstructed tracks associated with 
the hard-scatter vertex which are not in the hard term (collected in the soft term). In 
order to mitigate the effects of pile-up, the $\met$ is refined by using object-level 
corrections for the identified electrons, muons, jets and $\tau_{\text{had-vis}}$ 
candidates in the hard term. As the soft term contains only tracks associated with 
the hard-scatter vertex, it is robust against pile-up.

%% file: Selection.tex
Charged Higgs bosons are searched for in the topology 
$pp \to [b]t\Hp \to [b](jjb)(\tau_{\text{had}}\nu)$. Events 
collected using an $\met$ trigger with a threshold at 70~\GeV\ 
are considered. After ensuring that no jets are consistent 
with having originated from instrumental effects or non-collision 
background, each event is required to contain 
one $\tau_{\text{had-vis}}$ with $\pt^\tau > \unit[40]{\GeV}$ (only 
the highest-$\pt^\tau$ candidate must fulfil the identification 
criteria described in Section~\ref{sec:PhysicsObjects}), three 
or more jets with $\pt > \unit[25]{\GeV}$, of which at least one is 
$b$-tagged, no electron or muon with a transverse energy or momentum above 
20~\GeV, and to have $\met > \unit[150]{\GeV}$. For the selected events, 
the transverse mass $\mt$ of the $\tau_{\text{had-vis}}$ and 
$\met$ system is defined as:

\begin{equation}
  \label{eq:mt}
\mt=\sqrt{ 2 \pt^\tau \met (1-\cos \Delta\phi_{\tau,\text{miss}}) },
\end{equation}

where $\Delta\phi_{\tau,\text{miss}}$ is the azimuthal angle between the 
$\tau_{\text{had-vis}}$ and the direction of the missing transverse momentum. 
This discriminating variable takes values lower than the $W$ boson mass 
for $W\rightarrow\tau\nu$ decays in background events and lower than the 
$\Hp$ mass for signal events, in the absence of detector resolution effects. 
A requirement of $\mt > \unit[50]{\GeV}$ is applied in order to reject events with 
mismeasured $\met$, where $\tau_{\text{had-vis}}$ is nearly aligned with 
the direction of the missing transverse momentum.
 
The $\met$ trigger efficiency is measured in data and then used to 
reweight the simulated events, rather than relying on the $\met$ trigger 
in the simulated samples. This measurement
is performed in a control region of the data that is orthogonal to the signal 
region described above, while retaining as many similarities as possible. 
For this purpose, events passing a single-electron trigger with a transverse 
energy threshold at 24~\GeV\ are considered and required to contain exactly 
one electron matched to the corresponding trigger object, exactly one 
$\tau_{\text{had-vis}}$ and two or more jets, of which at least one 
is $b$-tagged. Both the electron and the $\tau_{\text{had-vis}}$ fulfil 
loose identification criteria in order to improve the statistical precision, 
with little impact on the measured $\met$ turn-on curve. 

\begin{figure}[ht]
\begin{center}
\includegraphics[width=0.8\textwidth]{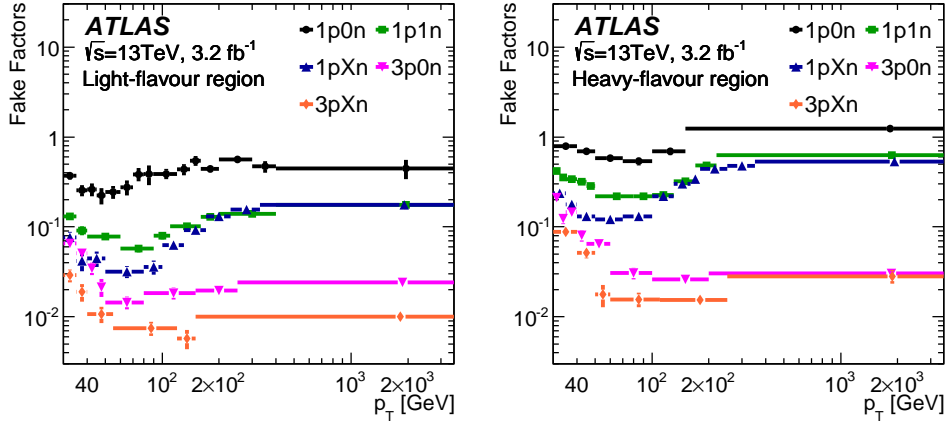}
\end{center}
\caption{Fake factors parameterised as a function of $\pt$ and the reconstructed 
$\tau_{\text{had}}$ decay mode, for a $b$-tagging score of the $\tau_{\text{had-vis}}$ 
candidate below (left, light-flavour region) or above (right, heavy-flavour region) the 
chosen threshold. The $\tau_{\text{had}}$ decay mode is referred to as 1p0n, 1p1n, 1pXn, 
3p0n or 3pXn, for a p-prong decay with n neutral particles $(\pi^{0})$, where $\mbox{X} \geq 2$ (1) 
for 1pXn (3pXn) decays. The threshold value for the $b$-tagging score is set to keep 
enough entries in each of the two bins, separately for each type of $\tau_{\text{had}}$ decay. 
The errors shown arise from the statistical uncertainty for a given $\pt$ bin.}
\label{FFs}
\end{figure}

The ``$\mbox{jet} \to \tau_{\text{had-vis}}$'' background includes multi-jet events and 
other processes where a quark- or gluon-initiated jet is reconstructed and 
selected as the $\tau_{\text{had-vis}}$ candidate. This background is estimated 
with a data-driven method that relies on the measurement of the rate at which 
jets are misidentified as $\tau_{\text{had-vis}}$, hereafter referred to as the 
fake factor (FF). For this purpose, a control region populated primarily with 
misidentified $\tau_{\text{had-vis}}$ candidates is defined by using the 
same requirements as for the signal region, except that $\met < \unit[80]{\GeV}$ and that 
the number of $b$-tagged jets is zero. The FF is defined as the ratio of the 
number of misidentified $\tau_{\text{had-vis}}$ candidates fulfilling the nominal 
object selection to the number of misidentified $\tau_{\text{had-vis}}$ candidates 
satisfying an ``anti-$\tau_{\text{had-vis}}$'' selection. This anti-$\tau_{\text{had-vis}}$ 
selection is defined by inverting the $\tau_{\text{had-vis}}$ identification criteria 
while maintaining a loose requirement on the BDT output score, which selects the 
same kind of objects mimicking $\tau_{\text{had-vis}}$ candidates as those fulfilling 
the identification criteria. In order to account for differences between gluon-, 
light-quark- and $b$-quark-initiated jets, FFs are parameterised as functions of 
$\pt$, the type of $\tau_{\text{had}}$ decay via the measured number of charged 
and neutral particles $(\pi^0)$~\cite{tau2}, and the $b$-tagging score, as illustrated in Figure~\ref{FFs}. 
For each type of $\tau_{\text{had}}$ decay, the threshold value for the $b$-tagging score of 
the $\tau_{\text{had-vis}}$ candidate is optimised to keep enough entries in each 
of the two bins, below and above the threshold. After correcting for $\tau_{\text{had-vis}}$ 
candidates not fulfilling the identification criteria but matching a true $\tau_{\text{had}}$ 
at generator level, the number of events with a misidentified $\tau_{\text{had-vis}}$ in the 
signal region ($N_\mathrm{fakes}^{\tau_{\text{had-vis}}}$) is derived from the subset of 
anti-$\tau_{\text{had-vis}}$ candidates as follows: 

\begin{equation} 
\label{eq:Nfakes}
N_\mathrm{fakes}^{\tau_{\text{had-vis}}} = \sum _i N_{\mathrm{anti}-\tau_{\text{had-vis}}}(i) \times \mbox{FF}(i),
\end{equation}
where the index $i$ refers to each bin in terms of $\pt$, type of $\tau_{\text{had}}$ decay 
and $b$-tagging score, in which the FF is evaluated.

Backgrounds arising from events in which an electron or muon is misidentified as
a $\tau_{\text{had-vis}}$ only contribute at the level of 5$\%$ to the 
total background, with misidentified muons contributing about one order of magnitude 
less than misidentified electrons. These backgrounds are estimated with simulation 
and include contributions from $\ttbar$, single-top-quark, $W/Z$+jets and diboson
processes. If an electron is misidentified as a $\tau_{\text{had-vis}}$, 
a correction factor is applied to the event in order to account for the misidentification 
rate measured in a $Z \to ee$ control region in data, where one electron is 
reconstructed as a $\tau_{\text{had-vis}}$. 
Charged leptons from in-flight decays in multi-jet events are accounted for in the
misidentified $\mbox{jet} \to \tau_{\text{had-vis}}$ background estimate.

\begin{figure}[h!]
  \begin{center}
    \includegraphics[width=0.57\textwidth]{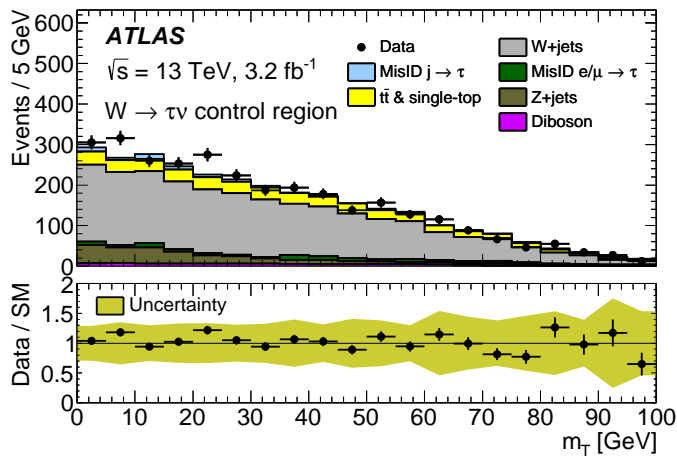}
  \end{center}
  \caption{Distribution of $\mt$ in the control region enriched with ${W \to \tau\nu}$ events, 
which differs from the nominal event selection by the requirements that $\mt < \unit[100]{\GeV}$ 
and that the number of $b$-tagged jets be zero. The ${W \to \tau\nu}$ 
background is normalised to the data through an overall scale factor. 
The total (statistical and systematic) uncertainties 
in the SM prediction are shown in the lower plot. }
  \label{CR-Wtaunu}
\end{figure}

\begin{figure}[h!]
  \begin{center}
    \includegraphics[width=0.57\textwidth]{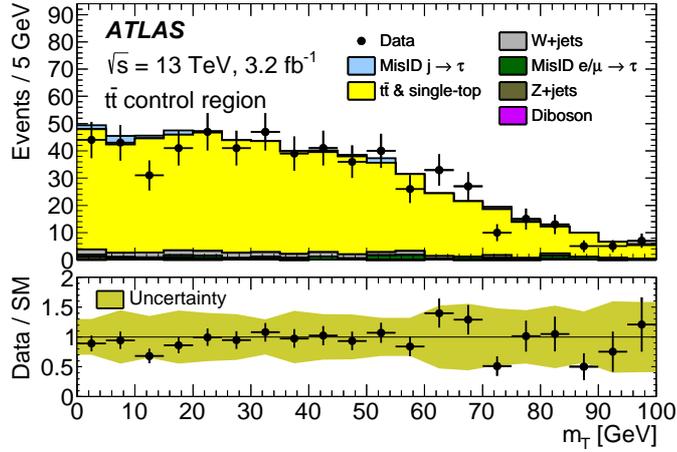}
  \end{center}
  \caption{Distribution of $\mt$ in the control region enriched with $t\bar{t}$ events, 
which differs from the nominal event selection by the requirements that $\mt < \unit[100]{\GeV}$ 
and that the number of $b$-tagged jets be at least two. The total (statistical and systematic) 
uncertainties in the SM prediction are shown in the lower plot.}
  \label{CR-ttbar}
\end{figure}

The backgrounds with a true $\tau_{\text{had}}$ are estimated using simulation. 
The two dominant processes, $t\bar{t}$ and $W \to \tau\nu$, are validated in two
dedicated control regions, which differ from the nominal event selection by the 
requirements that $\mt < \unit[100]{\GeV}$ (instead of $\mt > \unit[50]{\GeV}$) 
and that the number of $b$-tagged jets be either at least two (for the control 
region enriched with $t\bar{t}$ events) or zero (for the control region enriched 
with $W \to \tau\nu$ events). The latter is also used to correct the overall 
normalisation of the simulated $W \to \tau\nu$ background. The $\mt$ distributions 
that are predicted and measured in these two background-enriched control regions are 
displayed in Figures~\ref{CR-Wtaunu} and~\ref{CR-ttbar}. 
The relative signal contamination in the control region enriched in 
$W \to \tau\nu$ events is about two orders of magnitude smaller than the expected 
fraction of $\Hp \to \tau\nu$ events in the signal region. The control region enriched
in \ttbar~ events has a small overlap with the signal region, however the expected
signal contamination is negligible, about one order of magnitude smaller than the
expected fraction of $\Hp \to \tau\nu$ events in the signal region.


The expected number of background events in the signal region is shown in 
Table~\ref{tab:results_taujets_prefit}, together with the hypothetical contribution 
from charged Higgs bosons with a mass of 200 or 1000~\GeV, and with
$\sigma(pp \to [b]t\Hpm) \times \mbox{BR}(\Hpm \to \tau\nu)$ set to the prediction from the 
hMSSM scenario for $\tan\beta = 60$ (for a given mass, the expected signal yield increases 
quadratically with $\tan\beta$). The calculation of the production cross section is 
based on Refs.~\cite{xsec0,xsec1,xsec2,xsec3,xsec4}, while {\sc HDECAY}~\cite{Djouadi:1997yw} 
is used for computing the branching fraction. 
The signal acceptance at 200 (1000)~\GeV\ is 1.5\% (12\%), as evaluated with respect to 
simulated samples where both the $\tau$-lepton and the associated top quark decay 
inclusively. The event yield observed in $\unit[3.2]{\ifb}$ of data is also shown in
Table~\ref{tab:results_taujets_prefit} and found to be consistent with the expectation 
for the background-only hypothesis.

\begin{table}[h!]
\begin{center}
\begin{tabular}{l|l}
Sample & Event yield \\
\hline 
True $\tau_{\text{had}}$ & \\
 \hspace*{5mm}$t\bar{t}$ \& single-top-quark  & $590\phantom{.0}\pm170\phantom{.0}$ \\
 \hspace*{5mm}$W \to \tau\nu$  & $\phantom{0}58\phantom{.0}\pm\phantom{0}14\phantom{.0}$ \\
 \hspace*{5mm}$Z \to \tau\tau$  & $\phantom{00}6.4\pm\phantom{00}2.0$ \\
 \hspace*{5mm}diboson ($WW, WZ, ZZ$) & $\phantom{00}4.3\pm\phantom{00}1.3$ \\
Misidentified $e,\,\mu \to \tau_{\text{had-vis}}$   & $\phantom{0}40\phantom{.0}\pm\phantom{00}6\phantom{.1}$ \\
Misidentified $\mbox{jet} \to \tau_{\text{had-vis}}$ & $196\phantom{.0}\pm\phantom{0}24\phantom{.0}$ \\
\hline
All backgrounds   & $900\phantom{.0}\pm170\phantom{.0}$ \\
\hline
$H^+$ (200~\GeV), hMSSM $\tan\beta=60$ & $175\phantom{.0}\pm\phantom{0}28\phantom{.0}$ \\
$H^+$ (1000~\GeV), hMSSM $\tan\beta=60$ & $\phantom{00}2.0\pm\phantom{00}0.2$ \\
\hline
Data         &  $890$   \\
\hline
\end{tabular}
\caption{\label{tab:results_taujets_prefit} Expected event yields for the backgrounds and a hypothetical 
$\Hp$ signal after all selection criteria, and comparison with $\unit[3.2]{\ifb}$ of data. The values shown 
for the signal assume a charged Higgs boson mass of 200 or 1000~\GeV, with a cross section times 
branching fraction $\sigma(pp \to [b]t\Hpm) \times \mbox{BR}(\Hpm \to \tau\nu)$ corresponding to 
$\tan\beta=60$ in the hMSSM benchmark scenario. The uncertainties include statistical and systematic 
components.}
\end{center}
\end{table}

%% file: Systematics.tex
Several sources of systematic uncertainty, affecting the normalisation of signal 
and background processes or the shape of their distributions, are considered. 
The individual sources of systematic uncertainty are assumed to be uncorrelated. 
However, when applicable, correlations of a given systematic uncertainty are 
maintained across all processes. All systematic uncertainties are symmetrised 
with respect to the nominal value.

In order to assess the impact of most detector-related systematic uncertainties, in particular 
those arising from the simulation of pile-up and object reconstruction, the event selection 
is re-applied after shifting a particular parameter to its $\pm 1$ standard-deviation value. 
All instrumental systematic uncertainties arising from the reconstruction, identification 
and energy scale of electrons, muons, ($b$-tagged) jets and $\tau_{\text{had-vis}}$ candidates 
are considered. They are propagated to the reconstructed $\met$ and an additional uncertainty 
in its soft term is taken into account. The dominant detector-related systematic uncertainties for this 
search arise from the reconstruction and identification of $\tau_{\text{had-vis}}$ candidates, 
from the jet energy scale, from the $\tau_{\text{had-vis}}$ energy scale and from the $b$-tagging 
efficiency. Their impacts on the predicted event yield for the dominant background process 
($t\bar{t}$) are, respectively, 12\%, 11\%, 3\% and 2\%. 
Systematic uncertainties arising from the reconstruction, identification and energy scale 
of electrons and muons are found to be negligible. The luminosity uncertainty of 5\% is 
applied directly to the event yields of all simulated events.

The efficiency of the $\met$ trigger is measured in a control region of the data, as described in Section~\ref{sec:Analysis}. The 
parameterisation of the efficiency shows a small dependence on the identification criteria 
(loose versus nominal) for the electron and the $\tau_{\text{had-vis}}$ candidate, as well 
as on the minimum number of jets chosen for the control region. This results in small 
variations of the measured fit function. These variations, as well as the limited 
statistical precision of the bins used for the fit function and the resulting 
parameterisation, are accounted for as systematic uncertainties. In the signal 
region, the total systematic uncertainty arising from the $\met$ trigger 
efficiency measurement is about 2\%.

In the estimation of backgrounds with jets misidentified as $\tau_{\text{had-vis}}$, 
the dominant systematic uncertainties arise from the level of contamination of 
$\tau_{\text{had-vis}}$ objects matching a true $\tau_{\text{had}}$ decay at 
generator level and fulfilling the anti-$\tau_\text{had-vis}$ selection (varied 
by 50\%), from the statistical limitation due to the control sample size and from 
the requirement on the BDT score in the anti-$\tau_{\text{had-vis}}$ control sample. 
When changing the latter, different fractions of gluon- and quark-initiated jets populate the 
anti-$\tau_\text{had-vis}$ control region. The event topology (in particular the shape of 
the $\met$ and $\Delta\phi_{\tau,\text{miss}}$ distributions) also depends on the requirement 
imposed on the BDT score. The corresponding systematic uncertainty is assessed by considering 
the shape of the $\mt$ distribution obtained for two alternative cuts on the BDT score, which 
are symmetric around the nominal cut value. 
The impacts of the three systematic uncertainties listed above on the event yield of the 
background with jets misidentified as $\tau_{\text{had-vis}}$ are, respectively, 8\%, 6\% 
and 2\%.

The dominant background process with a true $\tau_{\text{had}}$ is the production 
of $t\bar{t}$ pairs and single-top-quark events, for which an overall cross-section 
uncertainty of 6\% is applied, incorporating scale, PDF+$\alpha_{\rm{s}}$ and top-quark 
mass uncertainties~\cite{Czakon:2011xx,Aliev:2010zk,Kant:2014oha}. In addition, 
systematic uncertainties due to the choice of parton shower and hadronisation models 
are derived by comparing $t\bar{t}$ events generated with {\sc Powheg-Box} interfaced 
to either {\sc Pythia}~v8.210 or {\sc Herwig++}~v2.7.1~\cite{herwigpp}, which uses the 
UEEE5~\cite{u3e5} underlying-event tune. The systematic uncertainties arising from 
initial- and final-state parton radiation, which modify the jet production rate, are 
computed with the same packages as for the baseline $t\bar{t}$ event generation, by 
setting the corresponding parameters in {\sc Pythia} to a range of values not excluded 
by the experimental data. Finally, the uncertainty due to the choice of matrix-element 
generator is evaluated by comparing samples generated with {\sc MadGraph5\_aMC@NLO} or {\sc Powheg-Box}, 
both using the CTEQ6L1 PDF set and interfaced to {\sc Herwig++}.
The impacts of the three systematic uncertainties listed above on the event yield of the $t\bar{t}$ 
background are, respectively, 16\%, 7\% and 15\%.

For the sub-leading background process with a true $\tau_{\text{had}}$, 
$W \to \tau\nu$, a systematic uncertainty of 3\% is assigned to the overall 
renormalisation factor, as obtained by changing various selection criteria for 
the control region enriched with such background events. For $Z$+jets and 
diboson production, theoretical uncertainties of 5\% and 6\% are considered, 
respectively, combining PDF+$\alpha_{\rm{s}}$ and scale variation uncertainties in 
quadrature.

Systematic uncertainties in the $\Hp$ signal generation are estimated as follows. 
The uncertainty arising from the QCD scale is obtained by varying the factorisation 
and renormalisation scale up and down by a factor of two. The largest variation of 
the signal acceptance is then symmetrised and taken as the scale uncertainty, 4--8\% 
depending on the $\Hp$ mass hypothesis. The variation of the signal acceptance 
with various PDF sets is estimated using {\sc LHAPDF}~\cite{lhapdf}, but is 
found to be negligible for all signal samples. Finally, the impact of A14 tune 
variations on the signal acceptance is estimated by adding in quadrature the 
positive and negative excursions from a subset of tune variations that cover 
underlying-event and jet-structure effects, as well as different aspects of extra 
jet production. This uncertainty amounts to 8--10\% and is of the same order as the 
sum in quadrature of the detector-related systematic uncertainties for the $\Hp$ signal 
samples.

%% file: Results.tex
In order to test the compatibility of the data with the background-only and 
signal+background hypotheses, a profile likelihood ratio~\cite{combi2011} is 
used, with $\mt$ as the discriminating variable. The statistical analysis is 
based on a binned likelihood function for this distribution. All systematic 
uncertainties from theoretical or experimental sources are implemented as 
nuisance parameters. The parameter of interest (or signal strength) 
$\mu \equiv \sigma(pp \to [b]t\Hpm) \times \mbox{BR}(\Hpm \to \tau\nu)$, 
and the nuisance parameters are simultaneously fitted by means 
of a negative log-likelihood minimisation. 
Expected limits are derived using the asymptotic approximation of the distribution of 
the test statistic~\cite{Asimov}. 

\begin{figure}[htp]
  \centering
  \includegraphics[width=0.65\textwidth]{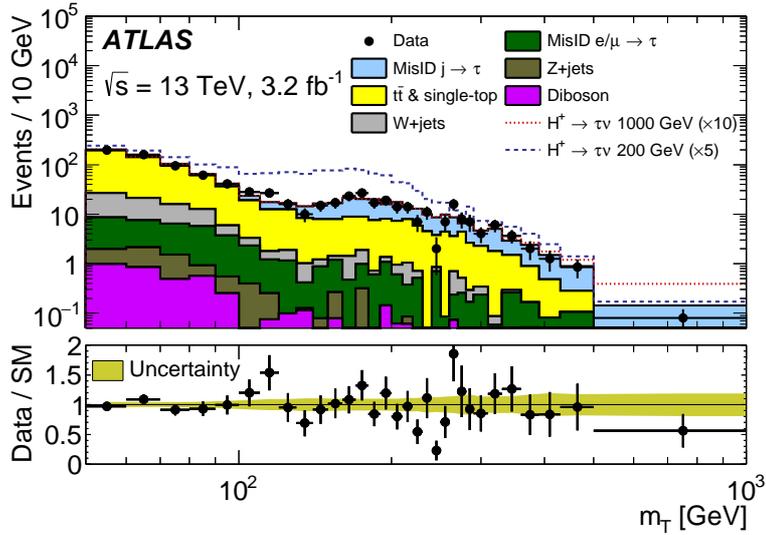}
  \caption{
    Distribution of $\mt$ after full event selection and a fit to the data with the background-only 
hypothesis. The horizontal axis starts at $\mt = \unit[50]{\GeV}$ and is in logarithmic scale. Two 
$\Hp$ signal hypotheses are included separately on the stack. The signal sample at 200 (1000)~\GeV\ 
is scaled to 5 (10) times the cross section predicted at $\tan\beta = 60$ in the hMSSM benchmark scenario. 
Bins are 10~\GeV\ in width up to 310~\GeV\ and then have a varying size.  The last bin includes all overflow events. 
The total (statistical and systematic) uncertainties in the SM prediction are shown in the lower plot. 
  }
  \label{fig:post_fit}
\end{figure}

\begin{figure}[htp]
  \centering
  \includegraphics[width=0.65\textwidth]{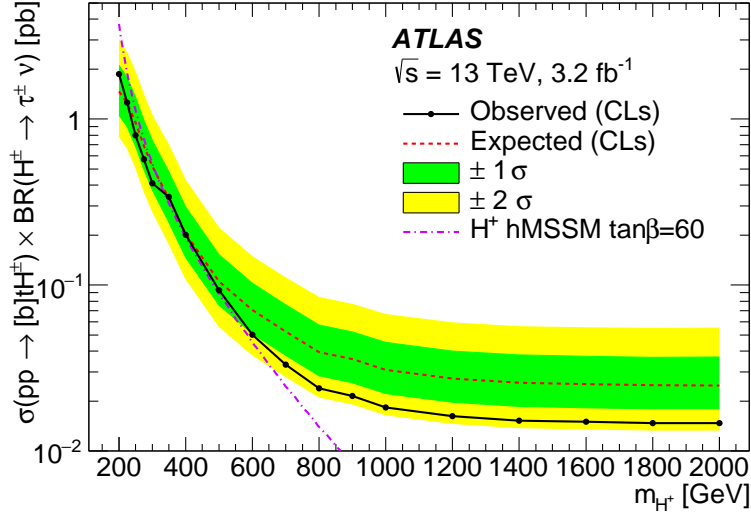}
  \caption{
    Observed and expected 95\% CL exclusion limits for heavy charged Higgs boson production 
as a function of $m_{\Hp}$ in 3.2~\invfb of $pp$ collision data. The prediction for  
$\sigma(pp \to [b]t\Hpm) \times \mbox{BR}(\Hpm \to \tau\nu)$ as a function of the 
charged Higgs boson mass is also shown as a dotted-dashed line, for $\tan\beta = 60$ 
in the hMSSM benchmark scenario. 
}
  \label{fig:limit_obs}
\end{figure}

Figure~\ref{fig:post_fit} shows the $\mt$ distribution obtained after a fit with the 
background-only hypothesis, together with the $\mt$ distributions corresponding to 
two $\Hp$ mass hypotheses: 200 and 1000~\GeV. The binning shown in Figure~\ref{fig:post_fit} 
is also used in the statistical analysis.  
The SM predictions are found to be consistent 
with the data, and exclusion limits are set on 
$\sigma(pp \to [b]t\Hpm) \times \mbox{BR}(\Hpm \to \tau\nu)$ 
by rejecting the signal hypothesis at the 95\% confidence level (CL) using the CL$_s$ 
procedure~\cite{CLs}. Figure~\ref{fig:limit_obs} shows the observed and expected 
exclusion limits. They agree within the uncertainties over the 
explored $\Hp$ mass range. The observed limits range from 1.9~pb to 15~fb in 
the mass range 200--2000~\GeV. For the largest charged Higgs boson mass hypotheses, 
the exclusion limits show very little dependence on $m_{\Hp}$, as there is only one 
bin entering the fit for $\mt > \unit[500]{\GeV}$. 
The impact from the various sources of systematic uncertainty on the  
expected 95\% CL exclusion limits are summarised in Table~\ref{tab:syst_taujets_postfit}, 
for $\Hp$ mass hypotheses of 200 and 1000~\GeV. 
The impact of the systematic uncertainties reported in Section~\ref{sec:syst} only represents 
the relative change in event yields. In the limit-setting procedure, however, $\mt$ shape 
variations are also taken into account, leading to a different relative importance of the 
various systematic uncertainties. 
Those with a large impact over the explored mass range are the $\tau_{\text{had-vis}}$ 
identification and energy-scale uncertainties, the $t\bar{t}$ background modelling 
uncertainties, and the statistical precision in the estimation of the background 
with a jet misidentified as a $\tau_{\text{had-vis}}$. For the larger $H^+$ mass 
hypotheses, the signal modelling uncertainties also have a significant impact.
The total uncertainty is dominated by the statistical uncertainty.

\begin{figure}[h!]
  \centering
  \includegraphics[width=0.5\textwidth]{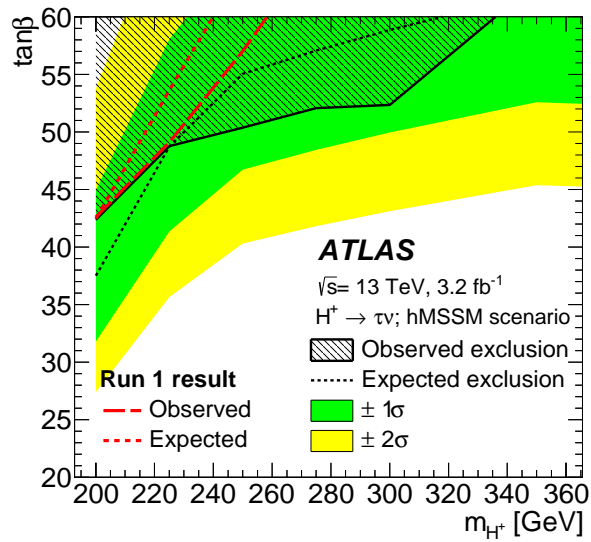}
  \caption{95\% CL exclusion limits on $\tan\beta$ as a function of $m_{\Hplus}$,
    shown in the context of the hMSSM, for the regions in which reliable theoretical 
    calculations exist ($\tan\beta \leq 60)$. As a comparison, the red curves show 
    the observed and expected exclusion limits from Run 1 analyses of $pp$ collisions 
    measured at $\rts = \unit[8]{\TeV}$ by ATLAS~\cite{taunupaper}.}
\label{plot-hmssm}
\end{figure}

The limits in Figure~\ref{fig:limit_obs} are presented together with an illustrative
signal prediction in the hMSSM benchmark scenario.
Figure~\ref{plot-hmssm} shows the 95\% CL exclusion limits on $\tan\beta$ 
as a function of $m_{\Hp}$ in the context of the hMSSM, compared with the 
Run 1 results. Values of $\tan\beta$ in the range 42--60 are excluded for
a charged Higgs boson mass of 200~\GeV. At $\tan\beta = 60$, above which no
reliable theoretical calculations exist, the $\Hp$ mass range from 200 to 340~\GeV\
is excluded. The limits of this search surpass those obtained with the $pp$ 
collision data at $\rts = \unit[8]{\TeV}$~\cite{taunupaper}.

\begin{table}[h!]
\begin{center}
\begin{tabular}{l|cc}
Source of systematic & \multicolumn{2}{c}{Impact on the expected limit (in \%)} \\
uncertainty & $m_{H^+} = 200~\GeV$ & $m_{H^+} = 1000~\GeV$ \\
\hline
Experimental & & \\
\hspace*{5mm} luminosity         &  $\phantom{0}2.0$     &  $\phantom{0}1.1$ \\
\hspace*{5mm} trigger             &  $<0.1$               &  $<0.1$ \\
\hspace*{5mm} $\tau_{\text{had-vis}}$                      &  $\phantom{0}2.7$           &  $\phantom{0}1.1$ \\
\hspace*{5mm} jet                                           &  $\phantom{0}0.4$           &  $<0.1$ \\
\hspace*{5mm} $\met$                                         &  $\phantom{0}0.3$           &  $<0.1$ \\
\hline
Fake factors & & \\
\hspace*{5mm} statistical limitation     &  $\phantom{0}4.5$     &  $\phantom{0}0.7$ \\
\hspace*{5mm} true $\tau_{\text{had}}$ contamination              &  $<0.1$           &  $<0.1$ \\
\hspace*{5mm} anti-$\tau_{\text{had-vis}}$ BDT score               &  $\phantom{0}0.2$    &  $\phantom{0}0.6$ \\
\hline
Signal and background models & & \\
\hspace*{5mm} $t\bar{t}$ cross section   &  $\phantom{0}0.2$     &  $<0.1$ \\
\hspace*{5mm} $t\bar{t}$ modelling        &  $\phantom{0}7.5$     &  $\phantom{0}1.0$ \\
\hspace*{5mm} $\Hplus$ signal modelling       &  $\phantom{0}1.4$     &  $\phantom{0}1.3$ \\

\hline
\end{tabular}
\caption{\label{tab:syst_taujets_postfit} Impact of various sources of uncertainty on 
the expected 95\% CL exclusion limit, for two $H^+$ mass hypotheses: 200 and 1000~\GeV. The impact
is obtained by comparing the nominal expected limit with the expected limit when
a certain set of uncertainties is not included in the limit-setting procedure.
}
\end{center}
\end{table}

%% file: conclusion.tex
A search for charged Higgs bosons produced in association with a single top quark and 
subsequently decaying via $\Hp \to \tau \nu$ is performed, based on fully hadronic final 
states. The dataset used for this analysis contains 3.2~\invfb\ of $pp$ collisions at 
$\rts = \unit[13]{\TeV}$, recorded in 2015 by the ATLAS detector at the LHC. The 
background-only hypothesis is found to be in agreement with the data. Upper limits are 
set on the production cross section times branching fraction between 1.9~pb and 15~fb for a 
charged Higgs boson mass range of 200--2000~\GeV. 
In the context of the hMSSM, values of $\tan\beta$ in the range 42--60 are excluded for
a charged Higgs boson mass of 200~\GeV. At $\tan\beta = 60$, above which no reliable 
theoretical calculations exist, the $\Hp$ mass range from 200 to 340~\GeV\
is excluded.

%% file: Acknowledgements.tex

We thank CERN for the very successful operation of the LHC, as well as the
support staff from our institutions without whom ATLAS could not be
operated efficiently.

We acknowledge the support of ANPCyT, Argentina; YerPhI, Armenia; ARC, Australia; BMWFW and FWF, Austria; ANAS, Azerbaijan; SSTC, Belarus; CNPq and FAPESP, Brazil; NSERC, NRC and CFI, Canada; CERN; CONICYT, Chile; CAS, MOST and NSFC, China; COLCIENCIAS, Colombia; MSMT CR, MPO CR and VSC CR, Czech Republic; DNRF and DNSRC, Denmark; IN2P3-CNRS, CEA-DSM/IRFU, France; GNSF, Georgia; BMBF, HGF, and MPG, Germany; GSRT, Greece; RGC, Hong Kong SAR, China; ISF, I-CORE and Benoziyo Center, Israel; INFN, Italy; MEXT and JSPS, Japan; CNRST, Morocco; FOM and NWO, Netherlands; RCN, Norway; MNiSW and NCN, Poland; FCT, Portugal; MNE/IFA, Romania; MES of Russia and NRC KI, Russian Federation; JINR; MESTD, Serbia; MSSR, Slovakia; ARRS and MIZ\v{S}, Slovenia; DST/NRF, South Africa; MINECO, Spain; SRC and Wallenberg Foundation, Sweden; SERI, SNSF and Cantons of Bern and Geneva, Switzerland; MOST, Taiwan; TAEK, Turkey; STFC, United Kingdom; DOE and NSF, United States of America. In addition, individual groups and members have received support from BCKDF, the Canada Council, CANARIE, CRC, Compute Canada, FQRNT, and the Ontario Innovation Trust, Canada; EPLANET, ERC, FP7, Horizon 2020 and Marie Sk{\l}odowska-Curie Actions, European Union; Investissements d'Avenir Labex and Idex, ANR, R{\'e}gion Auvergne and Fondation Partager le Savoir, France; DFG and AvH Foundation, Germany; Herakleitos, Thales and Aristeia programmes co-financed by EU-ESF and the Greek NSRF; BSF, GIF and Minerva, Israel; BRF, Norway; Generalitat de Catalunya, Generalitat Valenciana, Spain; the Royal Society and Leverhulme Trust, United Kingdom.

The crucial computing support from all WLCG partners is acknowledged
gratefully, in particular from CERN and the ATLAS Tier-1 facilities at
TRIUMF (Canada), NDGF (Denmark, Norway, Sweden), CC-IN2P3 (France),
KIT/GridKA (Germany), INFN-CNAF (Italy), NL-T1 (Netherlands), PIC (Spain),
ASGC (Taiwan), RAL (UK) and BNL (USA) and in the Tier-2 facilities
worldwide.

%% file: atlas_authlist.tex

\begin{flushleft}
{\Large The ATLAS Collaboration}

\bigskip

M.~Aaboud$^\textrm{\scriptsize 135d}$,
G.~Aad$^\textrm{\scriptsize 86}$,
B.~Abbott$^\textrm{\scriptsize 113}$,
J.~Abdallah$^\textrm{\scriptsize 64}$,
O.~Abdinov$^\textrm{\scriptsize 11}$,
B.~Abeloos$^\textrm{\scriptsize 117}$,
R.~Aben$^\textrm{\scriptsize 107}$,
O.S.~AbouZeid$^\textrm{\scriptsize 137}$,
N.L.~Abraham$^\textrm{\scriptsize 149}$,
H.~Abramowicz$^\textrm{\scriptsize 153}$
% ...
\bigskip
\\
% ...
$^{11}$ Institute of Physics, Azerbaijan Academy of Sciences, Baku, Azerbaijan\\
% ...
$^{64}$ University of Iowa, Iowa City IA, United States of America\\
% ...
$^{86}$ CPPM, Aix-Marseille Universit{\'e} and CNRS/IN2P3, Marseille, France\\
% ...
$^{107}$ Nikhef National Institute for Subatomic Physics and University of Amsterdam, Amsterdam, Netherlands\\
% ...
$^{113}$ Homer L. Dodge Department of Physics and Astronomy, University of Oklahoma, Norman OK, United States of America\\
% ...
$^{117}$ LAL, Univ. Paris-Sud, CNRS/IN2P3, Universit{\'e} Paris-Saclay, Orsay, France\\
% ...
$^{135}$ $^{(a)}$ Facult{\'e} des Sciences Ain Chock, R{\'e}seau Universitaire de Physique des Hautes Energies - Universit{\'e} Hassan II, Casablanca; $^{(b)}$ Centre National de l'Energie des Sciences Techniques Nucleaires, Rabat; $^{(c)}$ Facult{\'e} des Sciences Semlalia, Universit{\'e} Cadi Ayyad, LPHEA-Marrakech; $^{(d)}$ Facult{\'e} des Sciences, Universit{\'e} Mohamed Premier and LPTPM, Oujda; $^{(e)}$ Facult{\'e} des sciences, Universit{\'e} Mohammed V, Rabat, Morocco\\
% ...
$^{137}$ Santa Cruz Institute for Particle Physics, University of California Santa Cruz, Santa Cruz CA, United States of America\\
% ...
$^{149}$ Department of Physics and Astronomy, University of Sussex, Brighton, United Kingdom\\
% ...
$^{153}$ Raymond and Beverly Sackler School of Physics and Astronomy, Tel Aviv University, Tel Aviv, Israel\\
% ...
$^{a}$ Also at Department of Physics, King's College London, London, United Kingdom\\
$^{b}$ Also at Institute of Physics, Azerbaijan Academy of Sciences, Baku, Azerbaijan\\
$^{c}$ Also at Novosibirsk State University, Novosibirsk, Russia\\
$^{d}$ Also at TRIUMF, Vancouver BC, Canada\\
$^{e}$ Also at Department of Physics {\&} Astronomy, University of Louisville, Louisville, KY, United States of America\\
% ...
\end{flushleft}